\documentclass[preprint,aps,tightenlines,floatfix,showpacs]{revtex4}

\usepackage{graphicx}
\usepackage{bm}
\usepackage{amsmath}
\usepackage{amssymb}

\begin{document}

\title{ A multi-channel model for an $\alpha$ plus ${}^6$He nucleus cluster}

 \author{K. Amos$^{(1,4)}$}
  \email{amos@unimelb.edu.au}
  \author{L. Canton$^{(2)}$}
  \author{P. R. Fraser$^{(3)}$}
  \author{\mbox{S. Karataglidis$^{(1,4)}$}}
  \author{J. P. Svenne$^{(5)}$}
  \author{D. van der Knijff$^{(1)}$}

  \affiliation{$^{(1)}$ School  of Physics,  University of  Melbourne,
    Victoria 3010, Australia}
  \affiliation{$^{(2)}$ Istituto  Nazionale  di  Fisica  Nucleare,
    Sezione di Padova, Padova I-35131, Italia}
  \affiliation{$^{(3)}$ Department of Physics, Astronomy and Medical Radiation Sciences,
   Curtin University, GPO Box U1987, Perth 6845, Australia}
  \affiliation{$^{(4)}$ Department of Physics, University of Johannesburg,
    P.O. Box 524 Auckland Park, 2006, South Africa}
  \affiliation{$^{(5)}$ Department  of  Physics  and Astronomy,
    University of Manitoba, and Winnipeg Institute for Theoretical Physics,
    Winnipeg, Manitoba, Canada R3T 2N2}


\date{\today}

\begin{abstract}
A multi-channel algebraic scattering (MCAS) method has been used to
solve coupled sets of Lippmann-Schwinger equations for the
$\alpha$+${}^6$He cluster system, so finding a model spectrum for
${}^{10}$Be to more than 10~MeV excitation.  Three states of ${}^6$He
are included and the resonance character of the two excited states
taken into account in finding solutions.  A model Hamiltonian has been
found that gives very good agreement with the known bound states and
with some low-lying resonances of ${}^{10}$Be. More resonance states
are predicted than have as yet been observed.  The method also yields
$S$-matrices which we have used to evaluate low-energy
${}^6$He-$\alpha$ scattering cross sections. Reasonable reproduction
of low-energy differential cross sections and of energy variation of
cross sections measured at fixed scattering angles is found.
\end{abstract}

\maketitle

\section{Introduction}

Nuclear reactions can and have been used as tests of models of nuclear
structure, through analyses of data therefrom.  In the low-energy
regime, they are the primary energy-generation mechanisms in
stars. However, in a stellar environ, those reactions typically take
place at energies much lower than those accessible by laboratory
experiments.  For example, at stellar energies ($\le \sim 300$ keV),
radiative capture rates are usually too small to be measured directly.
Thus, estimates of reaction cross sections important in astrophysics
application have to be extrapolated from data measured at higher
energies.  Theoretical extrapolations are difficult due to the size of
uncertainties that accompany the lowest-energy results measured, as
well as from any specific influence of resonances in the Gamow window
regime.  Nonetheless, some cases are suited to direct theoretical
analysis by using few-body techniques.  Three- and four-body systems
can be treated by finding solutions of the Faddeev~\cite{Fa61} and
Faddeev-Yakubovsky~\cite{Ya67} equations, respectively, with the later
in the Alt-Grassberger-Sandhas form~\cite{Al72} and using realistic
nucleon-nucleon interactions. However, calculations of scattering with
these methods are technically complex.

There have been many attempts to theoretically establish microscopic
models of both nuclear structure and reactions (see
~\cite{Du11,Ro16,Na16} for many references).  In~\cite{Ro16}, one such
theory was developed to specify the nuclear optical potential from
first principles.  An objective was to have a well established
potential for use in evaluation of capture rates, especially of
neutron capture rates on rare isotopes of nuclei; rates that are
needed in simulations of cataclysmic events such as supernovae
explosions.  But, it has been noted~\cite{Ro16} that only selected
nuclei, and specific reaction channels, can be addressed with the
various {\it ab-initio} methods so far developed.

Thus, it remains a standard approach to use phenomenological optical
potentials between nuclear clusters, from which relative motion
functions can be used to give capture cross sections at energies in
the Gamow window.  In most studies, local forms of phenomenological
optical potentials have been used for simplicity.  With stable nuclei,
there exist global parameterisations obtained from fits to much data.
However, the application of global parameterization parameter values
to exotic nuclear systems is unreliable.  Furthermore, optical
potentials are known to be non-local due to the character of the
underlying in-medium nucleon-nucleon interactions, the Pauli principle
that results in exchange scattering amplitudes, and specific
coupled-channel effects.  These features have been
described~\cite{Am00}, and at low energies, the role of
coupled-channels has been shown~\cite{Fr08a} to be more complex than
the approximate treatment of nonlocality via the Perey
effect~\cite{Pe62}.

However, if coupled-channel and Pauli principle effects can be taken
into account, local phenomenological nuclear interactions remain
useful for any simplicity they give in evaluations. As an example, a
phenomenological two-cluster model with a local two-nucleus
interaction~\cite{Du11} has been used to describe low energy
scattering and properties of the compound system bound states.  The
Pauli principle was taken into account in the multi-nucleon systems by
considering group representations of the ground states of individual
elements of the cluster as well as of the compound.  Using a sum of
Gaussians for the two-nuclei interactions, fairly good descriptions of
low energy nucleon-nucleus and $\alpha$-nucleus scattering data were
found. Of note, the potential parameters were partial wave dependent
so that for the $\alpha$ and light nuclei systems, $s,p,d$- and
$f$-partial waves contribute to the cross sections, even for energies
of few MeV.

In this paper, we report on studies made in a similar vein, using a
multi-channel algebraic scattering method (MCAS) to calculate low
energy scattering of two nuclei and the low excitation spectrum of
their compound (bound and resonant). Local forms are taken for the
interactions for the coupled-channel problem and allowance for the
effects of the Pauli principle has been made.  In particular we report
results of an application of the MCAS method to consider ${}^{10}$Be
as a coupled-channel problem of an $\alpha$-${}^6$He cluster.  We
consider the coupling to be with three low-excitation states (the
ground and two resonance states) of $^6$He.  Besides giving a spectrum
for ${}^{10}$Be, the MCAS programs also have been used to predict
low-energy cross sections for ${}^6$He-$\alpha$ elastic scattering.
Results are compared to a number of the measured data sets~\cite{Su13}
taken for a small range of energies and for various scattering angles.

The experimental results from the low-energy scattering of ${}^6$He
ions off $\alpha$ particles~\cite{Su13} are of great interest since
the ${}^6$He nucleus is weakly bound and is deemed to have an extended
neutron distribution that has been termed a `halo'. The $\alpha$
particle, in contrast, is strongly bound with its first excited state
at over 20~MeV excitation.  The compound system, ${}^{10}$Be, is also
quite strongly bound with the $\alpha$+${}^6$He threshold lying at
7.413~MeV in the spectrum.  Given that the $\alpha$+$\alpha$ compound,
${}^8$Be, is unbound, the two additional neutrons weakly bound in
${}^6$He are attached in covalent bounding orbits in the cluster of
that nucleus with an $\alpha$ to form ${}^{10}$Be.

The nucleus ${}^6$He lies close to the neutron drip line. Its ground
state lies 0.973~MeV below the two neutron break-up threshold and that
state $\beta^-$ decays.  The next two states are resonances with
energy centroids of 1.797 and 5.6~MeV~\cite{Ti02} and they decay by
two neutron (equivalently $\alpha$) emissions. The first excited
resonance state is narrow (113 keV) while the second is quite broad
(12~MeV~\cite{Ti02}).  Such properties have had impact in cluster
model evaluations of spectra of other compound
systems~\cite{Fr08,Ca11,Fr13,Fr16a}.

In the following section, we specify the model we have used to define
the matrix of potentials for the coupled-channel problem.  Then in
Sec.~\ref{be10} we report the spectra found with it for ${}^{10}$Be
treated as an $\alpha$+${}^6$He cluster.  In Sec.~\ref{scatt}, we
present and discuss elastic scattering cross sections of ${}^6$He ions
scattering from an $\alpha$-particle at low c.m. energies. Conclusions
are given in Sec.~\ref{conclus}.

\section{The model for the $\alpha$-nucleus matrix of potentials}
\label{potmat}

The $\alpha$-${}^6$He matrix of potentials have been defined using a
rotational collective model for the interaction~\cite{Ta65}.  The
potential assumed in describing the cluster system is allowed to have
central ($V_0$), $\ell^2$-dependent ($V_{\ell \ell}$), and
orbit-nuclear spin ($V_{\ell I}$) components.  This potential is also
taken to have quadrupole deformation.

There is an additional interaction considered to account for pair
correlations that affect the energies of $0^+$ states ${}^{10}$Be.
BCS theory, developed for the description of superconductivity, was
found very useful in treating the effect of pairing interactions in
even-mass nuclei, and delineating the consequences of correlations
induced by those interactions. In symbolic form, Hamiltonians have
been taken as
\begin{displaymath}
H = H_0 + H_{pair} + H_{Q \cdot Q},
\end{displaymath}
where $H_0$ is the part of the Hamiltonian that describes
single-particle motion in a self-consistent potential supplemented by
pairing ($H_{pair}$) and quadrupole-quadrupole ($H_{Q \cdot Q}$)
forces.  In Chapter 11 of Ref.~\cite{Ro10} on pairing force theory, it
is shown that a pairing interaction of monopole type leads to
additional binding in nuclear (ground) states of even mass nuclei, and
is considered to cause the even-odd $A$ mass difference.
Concomitantly, it can yield a large energy gap between the ground
$0^+$ and first excited states (usually a $2^+$).  Additionally, from
studies of collectivity in heavy nuclei~\cite{Iw76,Sa79}, the mutual
interplay between dynamical pairing and quadrupole correlations
explained anomalously low excitation energies of excited $0^+$ states.
With light nuclei, and ${}^{10}$Be in particular, a two-phonon pairing
vibrational ($2p-2h$) state is anticipated to lie at an energy of
4.8~MeV above the ground state~\cite{Ba99}; an energy which emerges
directly from two differences in binding energies, namely
[(BE(${}^{10}$Be) - BE(${}^8$Be)] - [BE(${}^{12}$Be) -
  BE(${}^{10}$Be)].  Pairing and correlation effects then are
relevant, particularly in defining the location of the $0^+$ states in
the spectrum of ${}^{10}$Be when treated as an $\alpha$+${}^6$He
cluster.  With the quite simple collective model form we use, those
additional properties are approximated by using a monopole potential
acting in all channels leading to $0^+$ states in the compound system.

To outline the MCAS method, consider a basis of channel states defined
by the coupling
\begin{equation}
\left| c \right\rangle = \left| \ell I J^\pi \right\rangle =
\biggl[
\left| \ell\right\rangle \otimes \left| \psi_I \right\rangle
\biggr]_J^{M,\pi}\ ,
\label{ch-state}
\end{equation}
where $\ell$ is the orbital angular momentum of relative motion of a
spin-0 projectile on the target whose states are $\left| \psi_I^{(N)}
\right\rangle$.  With each $J^\pi$ hereafter understood, and by
disregarding deformation temporarily, the ($\alpha$-nucleus) potential
matrices may be written
\begin{align}
V_{cc'}(r) = \langle \ell I  \left|\ W(r)\ \right|
\ell' I' \rangle
=& \bigg[ V_0 \delta_{c'c} f(r) + V_{\ell \ell} f(r) 
[ {\bf {\ell \cdot \ell}} ]  + V_{II} f(r) [{\bf I \cdot I}] 
+  V_{\ell I} g(r) [ {\bf {\ell \cdot I}} ]\bigg]_{cc'}
\nonumber\\
&\hspace*{0.5cm} + V_{mono} \delta_{c'c} \delta_{J^\pi = 0^+} f(r)\,,
\label{www1}
\end{align}
in which local form factors have been assumed.  Typically they are
specified as Woods-Saxon functions,
\begin{equation}
f(r) = \left[1 + e^{\left( \frac{r-R}{a} \right)} \right]^{-1}
\hspace*{0.3cm} ; \hspace*{0.3cm} g(r) = \frac{1}{r} \frac{df(r)}{dr} \,.
\label{radforms}
\end{equation}

Deformation then is included with the nuclear surface defined by
\begin{equation} 
R = R(\theta,\phi) = R_0 (1 + \epsilon), 
\end{equation}
where, for a rotational model of the target and in the space-fixed frame,
\begin{equation}
\epsilon = \sum_L \sqrt\frac{4\pi}{(2L+1)} \beta_L 
\left[ {\mathbf Y}_L(\Omega) \cdot  {\mathbf Y}_L(\zeta) \right]\,;
\end{equation}
$\beta_L$ are deformation parameters and $\zeta$ are the Euler angles
for the transformation from the body-fixed to space-fixed
frames. $\Omega = (\theta \phi)$ are the angles defining the surface
in the space-fixed frame.  Expanding $f(r-R(\theta\phi)) = f(r - R_0
(1 + \epsilon))$ to order $\epsilon^2$ gives
\begin{equation}
f(r) \to f_0(r) + \epsilon \left[\frac {df(r)}{d \epsilon}\right]_0
+ \frac {1}{2} \epsilon^2 \left[ \frac {d^2 f(r)}{d \epsilon^2}\right]_0
= f_0(r) - R_0 \frac{df_0(r)}{dr}\ \epsilon
+ \frac {1}{2} R_0^2\  \frac {d^2 f_0(r)}{d r^2}\ \epsilon^2 ,
\label{Eqn4}
\end{equation}
There is a similar equation for $g(r)$.

When collective models are used to specify the matrix of interaction
potentials acting between a nuclear projectile and a set of states of
a target nucleus, there are problems in satisfying the Pauli
principle~\cite{Ca05}.  In the MCAS method the effects of the Pauli
principle are met by inclusion of a set of orthogonalizing
pseudo-potentials (OPP)~\cite{Am03}; a technique that was developed in
studies of cluster physics~\cite{Kr74,Ku78} as a variant of the
Orthogonality Condition Model of Saito~\cite{Sa69}. It accounted for
the effects of Pauli blocking in the relative motion of two clusters
comprised of fermion constituents.  The OPP can also be used for the
situation with partially occupied levels being Pauli
hindered. Schmid~\cite{Sc78} notes that states can be Pauli-forbidden,
Pauli-allowed, or Pauli-suppressed; the last being what we have called
Pauli hindrance in previous applications of MCAS~\cite{Ca06,Am13}.

To orthogonalize states describing intra-cluster motion with respect
to the deeply-bound Pauli forbidden states, MCAS uses highly nonlocal
OPP terms embedded in a coupled-channel context.  The matrix of
interaction potentials (in coordinate space) to be used has the form
\begin{equation}
{\cal V}_{cc'} = V_{cc'}(r) \delta(r-r') + 
\lambda_c A_c(r) A_{c'}(r') \delta_{cc'} .
\end{equation}
$V_{cc'}(r)$ is the nuclear interaction potential and $\lambda_c$ is
the scale, in MeV, used to satisfy the Pauli principle. Pauli blocking
of the specific orbit in a particular channel, $c$, is achieved by
using a very large $\lambda_c$ value. That value should be infinite
but for all practical purposes $10^6$~MeV suffices.  Pauli allowed
states have $\lambda_c = 0$.  Pauli hindrance has values $0 <
\lambda_c < \infty\ (10^6)$.  The $A_c(r)$ are bound state wave
functions (of the $\alpha$ in this application) associated with the
diagonal nuclear interactions $V_{cc}(r)$ for the relevant orbital
angular momentum in each channel $c$.

\section{${}^{10}{\rm Be}$ as a coupled $\alpha$+${}^6{\rm He}$ system}
\label{be10}

This system is of interest given the (relatively) recent experimental
studies of the elastic scattering of ${}^6$He from an $\alpha$
target~\cite{Su13}, as well as of resonant $\alpha$ capture on
${}^6$He, which seek limits on clustering in states of the compound,
${}^{10}$Be.  We have used the coupled-channel Hamiltonian described
above to find as good a representation of the spectrum of ${}^{10}$Be
as possible, up to the first break-up threshold (of $n$+${}^9$Be) at
6.812~MeV and just beyond.  The $\alpha$+${}^6$He threshold lies at
7.413~MeV above the ground state.  To 6.812~MeV excitation there are
six known states, of which four have positive, and two negative,
parity. Two more states (one of each parity) lie at excitations close
to these thresholds and the rest have energy centroids greater than
9.2~MeV in the spectrum.  Thus, all low-excitation resonances in
${}^{10}$Be, save for the $3_1^-$ one at 7.31~MeV, may decay by
particle emission of a neutron and/or an $\alpha$-particle.

In the MCAS evaluations, the coupled-channel Hamiltonians were formed
assuming that there were three states of relevance in the spectrum of
${}^6$He.  They are the ground ($0^+$) that $\beta^-$ decays and the
two excited resonance states that decay by two neutron break-up. The
first resonance has a $2^+$ spin-parity while the second is
uncertain~\cite{Ti02}.  We choose it to have a $2^+$ spin-parity as
was the case when using MCAS to find a good spectrum for ${}^7$Li
treated as a $p$+${}^6$He cluster~\cite{Ca06a}.  The $2^+_1$ and
$2^+_2$ states, being particle-unstable resonances, have the assigned
shape of a Lorenzian multiplied by a projectile-energy dependant
scaling factor, which has the form of a Wigner distribution. This
eliminates erroneous threshold and sub-threshold behaviour inherent in
a pure Lorenzian form. As necessary with such widths, an
energy-dependant correction is added to the target-state centroids to
restore causality. Full details are available in Ref.~\cite{Fr16a}.
The ${}^6$He states' properties, along with the strengths of the OPP by
which the $\alpha$ particle orbits are hindered, are shown in
Table~\ref{He-targs}.
\begin{table}[h]
\begin{ruledtabular}
\caption{\label{He-targs} The states of ${}^6$He used in the
  coupled-channel evaluations.  With all states the $\alpha$ $s$-orbit
  was presumed blocked using $\lambda_{1s} = 10^6$ in the OPP. All
  energies are in units of MeV.}
\begin{tabular}{cccccc}
state & Centroid & Width & OPP ($\lambda_{1p}$) & OPP ($\lambda_{1d}$) & OPP
($\lambda_{2s}$)\\
\hline
$0^+_{\rm g.s.}$ & 0.000 & 0.00 & 10.915 & 0.0 & 3.3 \\
$2^+_1$ & 1.797 & 0.113 & 11.1\ \ & 4.7 & 0.75\\
$2^+_2$ & 5.60 & 12.0\ \ & \ 9.42\ \ & 0.0 & 0.0\\ 
\end{tabular}
\end{ruledtabular}
\end{table}

The coupled-channel matrix of potentials defining the Hamiltonian was
specified by a collective rotational model with the parameter set
listed in Table~\ref{He6-tab}.
\begin{table}[h]
\begin{ruledtabular}
\caption{\label{He6-tab} The potential parameters used for the
  interactions in the $\alpha$+${}^6$He system.  All strengths are in
  MeV and lengths are in fermi.}
\begin{tabular}{cccc}
Pot. strengths & Negative & Positive & Geometry \\
\hline
$V_0$    & $-$42.13 & $-$41.85 & $R_0 = 2.58$ \\
$V_{\ell \ell}$ & \ \ 1.14 & \ \ 1.14 & $a_0 = 0.7$ \\
$V_{\ell I}$ & \ \ 1.07 &\ \ 1.07  & $\beta_2 =  0.7$\\
$V_{II}$ & \ \ 1.4\  & \  0.4\  &                \\
$V_{mono}$ &         &  $-$7.72 &  \\
\hline
Coulomb (charge dist.) & & & \\
\hline
${}^6$He & $R_c = 1.3$ & ${}^4$He & $R_c = 1.008$ \\
         & $a_c = 0.4$ &          & $a_c = 0.327$ \\  
         & $w_c = 0.31$ &         & $w_c = 0.445$ \\ 
\end{tabular}
\end{ruledtabular}
\end{table}
Shown at the bottom of Table~\ref{He6-tab} are the parameter values of
a three parameter Fermi (3pF) model we have used to define the charge
distributions of ${}^{4,6}$He in generating the Coulomb potentials in
the Hamiltonian.  The 3pF charge distribution is of the form
\begin{equation}
\rho_{ch}(r) = \rho_0 \frac{1 + w_c \left( \frac{r}{R_c} \right)^2}
{1 + \exp\left(\frac{r - R_c}{a_c} \right)} \,.
\label{3pF-dist}
\end{equation}
For an $\alpha$ particle, the parameters were determined by a fit to
the electron scattering form factor at low momentum~\cite{Ja74}.  No
such data exist as yet for ${}^6$He, though analysis of the isotope
shift of spectral lines~\cite{Wa04} gives a root-mean-square (rms)
charge radius of 2.054 fm, and we select a 3pF parameter set
consistent with that value.  Details of our use of these charge
distributions are given in Appendix~\ref{append}.

\begin{figure}[ht]
\scalebox{0.7}{\includegraphics*{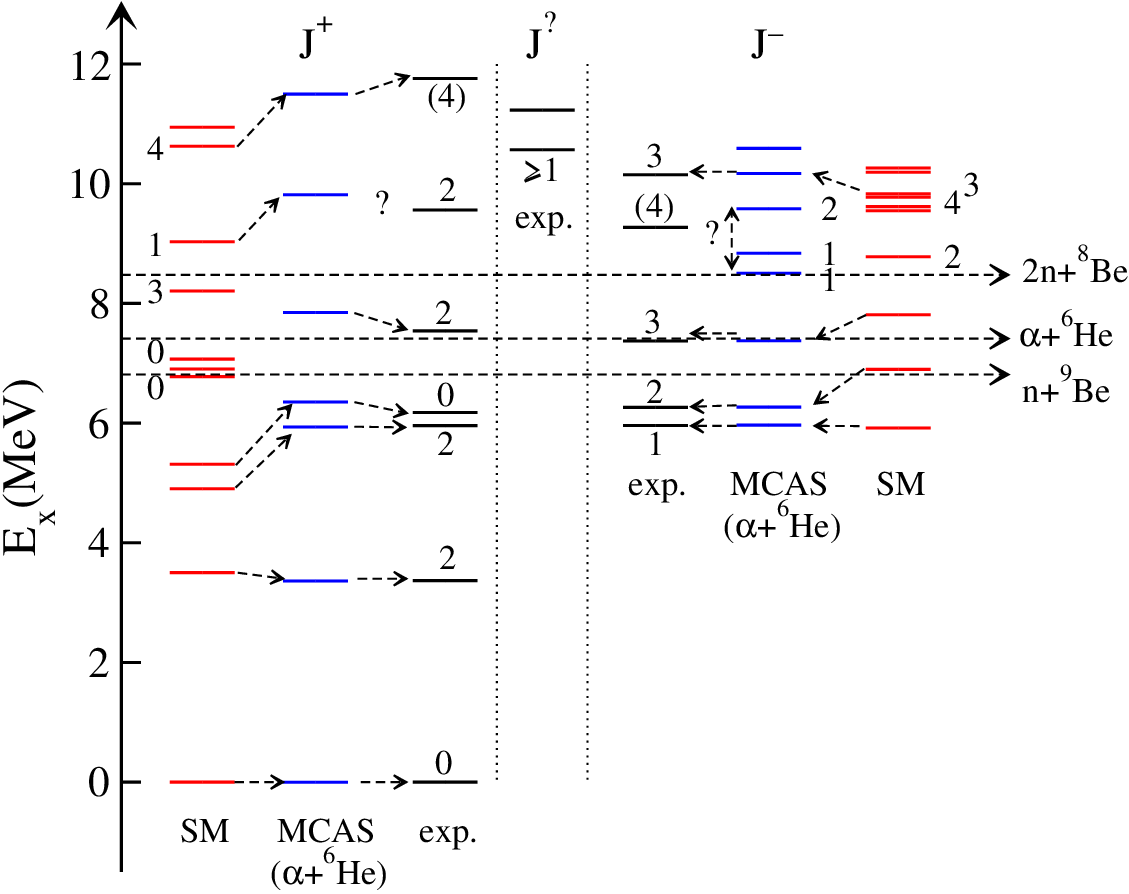}}
\caption{\label{Figure-1} (Color online) The low excitation spectra
  for ${}^{10}$Be found from the shell model calculations (left), from
  the data tabulation~\cite{Ti04} (center), and from the MCAS
  evaluations (left). Positive (negative) parity states are grouped on
  the left (right) of this figure.}
\end{figure}

Using MCAS for the $\alpha$+${}^6$He cluster gave the spectrum for
${}^{10}$Be identified with the label `MCAS' in Fig.~\ref{Figure-1}.
For clarity, we separate the positive and negative parity states in
the spectra on the left and right side of this figure.  The results
are compared with the known spectrum, labelled `exp.'.  Of note is
that, save for the uncertain assigned spin-parity of $(4^-)$, every
known state has a matching partner in the calculated spectrum with
excitation energies in quite good agreement.  For comparison, we have
found a spectrum for ${}^{10}$Be from shell model calculations made
using the WBT interaction~\cite{Wa92} with the OXBASH
program~\cite{Ox86}.  A no-core calculation was made by using a single
particle basis consisting of all single nucleon states from the
$0s$-shell up to, and including, the $(0h1f2p)$-shell. In this way,
the full six major shells have been taken into account.  Positive
parity states have been determined using the complete $(0+2+4)\hbar
\omega$ space while those for negative parity states were made in a
$(1+3)\hbar \omega$ space.  The results are shown in the columns
labelled `SM'.
In the shell model spectrum, there are additional positive parity
states, ($0^+_3, 3^+, 1^+$) states within 1.5~MeV above the $\alpha$
emission threshold, while there are additional negative parity states
($2^-_2, 0^-, 1^-_2, 1^-_3$) in the region to $\sim 10$~MeV
excitation.

Hence, both model results find reasonable to good agreement with the
known low-excitation spectrum; finding the eight known states to
within $\sim 1$~MeV of their listed energy values.  Both evaluations,
however, give a number of (unobserved) levels, especially in the
immediate region above the break-up thresholds.  Those thresholds
indicated on the diagram are for neutron ($n$+${}^9$Be; 6.812~MeV),
$\alpha$ ($\alpha$+${}^6$He; 7.413~MeV) and two-neutron
($2n$+${}^8$Be; 8.478~MeV) emissions.

Known properties of the low-excitation states in ${}^{10}$Be (energies
and widths) are also compared with the results of the MCAS evaluation
in Table~\ref{spectrum}. Here it must be noted that, as the known
resonances except for the $3^-_1$ state lie above the $n$+$^9$Be
threshold, the calculated widths are partial ones, and those for the
decay ${}^{10}$Be$\to n$+$^9$Be need be added to make a proper
comparison with the known values.
\begin{table}[h]
\begin{ruledtabular}
\caption{\label{spectrum} The known and MCAS evaluated spectrum of
  ${}^{10}$Be.  All energies are in units of MeV relative to the
  $\alpha$+${}^6$He threshold.}
\begin{tabular}{cccccc}
$J^\pi$ & E$_{Exp.}$ & $\Gamma_{Exp.}$ & $J^\pi$ & E$_{MCAS}$ & $\Gamma_{MCAS}$ \\
\hline
$0^+$ & {\bf $-$7.413} & & & {\bf $-$7.421} & \\
$2^+$ & {\bf $-$4.048} & & & {\bf $-$4.059} & \\
$2^+$ & {\bf $-$1.458} & & & {\bf $-$1.485} & \\
$1^-$ & {\bf $-$1.456} & & & {\bf $-$1.452} & \\    
$0^+$ & {\bf $-$1.237} & & & {\bf $-$1.065} & \\
$2^-$ & {\bf $-$1.153} & & & {\bf $-$1.153} & \\
$3^-$ & {\bf $-$0.045} & & &\ {\bf $-$0.042} &  \\
\hline
$2^+$ & {\bf \ 0.126} & 0.006 & & {\bf \ 0.429} & 0.006 \\
 & & & $1^-$ &\ 1.085 & 0.539 \\
$(4^-)$ & \ 1.854 & 0.154 & $1^-$ &\ 1.417 & 0.059 \\ 
$2^+$ & \  2.144 & 0.141 & $2^-$ & \ 2.164 & 0.130 \\    
 & & & $1^+$ & \ 2.395 & 0.277 \\
$3^-$ & \ {\bf 2.734} & 0.296 & & \ {\bf 2.710} & 0.119 \\
 $\ge 1$ & {\bf 3.157} & ? & $1^-$ & {\bf 3.172} & 0.998 \\
  ?   & 3.817 & 0.200 & & & \\
$4^+$ & {\bf 4.347} & 0.121 & & {\bf 4.077} & 5$\times 10^{-4}$\\
\end{tabular}
The question marks indicate that no value is given in the table~\cite{Ti04}.
\end{ruledtabular}
\end{table}

With MCAS, the match between the calculated and experimental spectra
${}^{10}$Be is very good up to $\sim 8$~MeV.  The calculated energies
of all (known) states in the region agree to within a few tens of keV,
and we emphasise those by bold type in the table.  The $3^-_2$
resonances has a matching MCAS results, with a centroid only 24~keV
from the established value and a well-matching centroid. The observed
$2^+_4$ state does not have a partner, with the same parity, though a
$2^-$ state is calculated with an energy difference of 20~keV and an
almost identical width.

To achieve these results with MCAS required 
\begin{itemize}
\item
A `pairing correlation effect' (monopole) components in the
$\alpha$-${}^6$He interaction.  This was needed to provide extra
binding for the $0^+_{1,2}$ states of the compound system
${}^{10}$Be. A shift of $\sim 2$~MeV from the values found with no
monopole term, was required.
\item
An OPP hindrance term for the $1p$-orbit of the $\alpha$ in each of
the three target states.  They only influence the results for the
negative parity states in ${}^{10}$Be, but were needed both to find
the correct sequence of spin-parities and the energy separations of
the three bound states.
\item
An OPP hindrance term for the $1d$-orbit of the $\alpha$ in the $^6$He
$2^+_1$ state.  This was necessary to increase the energy of the
$^{10}$Be $4^+$ resonance to near experiment. Without it, this state
had an energy of $\sim$7~MeV.
\item
The chosen width of the $2^+_2$ resonance state in ${}^6$He,
influences the positive parity spectrum of ${}^{10}$Be in the
resonance region.  However only when the value is quite small is there
noticeable variations in the spectrum.
\item
Varying the positive parity interaction strengths gave the following
effects:
\begin{enumerate}
\item
Setting $V^{(+)}_{II}=0$ changed the energies of all but that of the
$2^+_1$ by less than 0.4~MeV, and that of the ground not at all. The
$2^+_1$ state, however became much more bound.
\item
Setting $V^{(+)}_{\ell I}=0$ hardly affected the energies of the
ground ($0^+$) and $2^+_1$ states, but others varied by as much as
$\pm 1$~MeV.
\item
Setting $V^{(+)}_{\ell \ell} = 0$ did not affect the binding energy of
the ground state but changes those of all others noticeably, most
becoming much more bound.
\end{enumerate} 
\item
while, for the negative parity spectrum,
\begin{enumerate}
\item
Setting $V^{(-)}_{II}=0$ changed all but the binding of the $2^-_1$
state by less than $-0.3$~MeV. The $2^-_1$ state however was much more
bound.
\item
Setting $V^{(-)}_{\ell I} = 0$ gave only minor changes with states
being more bound by less than 0.2~MeV.
\item
Setting $V^{(-)}_{\ell \ell} = 0$ made most states more bound; an
effect equivalent to increasing the depth of the central component.
\item
The Pauli hindrance of the relative motion $p$-orbit was crucial in
finding the energies and splitting of the lowest two negative parity
states in particular.  Removing this hindrance lead to the $1^-_1$
state dropping in energy to become the ground state.
\end{enumerate}
\end{itemize}


\section{Scattering of ${}^6{\rm He}$ ions from $\alpha$ particles}
\label{scatt}

Data from elastic (and inelastic) scattering of a ${}^6$He radioactive
ion beam from $\alpha$-particles have been reported
recently~\cite{Su13} at low energies suitable for analysis using MCAS.
Data, both in the form of angular distributions at fixed energies and
differential energy cross sections for select angle values, have been
taken for a range of center of mass (c.m.) energies from $\sim$2 to
6~MeV. With respect to the spectrum of the compound ${}^{10}$Be, this
energy range (above threshold) coincides with an excitation energy
$\sim$9.4 to 13.4~MeV.

With the coupled-channel Hamiltonian defined to best represent the
known sub-threshold states in ${}^{10}$Be, numerous resonance states
are found with higher energies. The resonance states with well-known
$J^\pi$ in ${}^{10}$Be lying above the $\alpha$+${}^6$He threshold
(7.41~MeV), and with centroid energies defined with respect to that
threshold, are the $2_3^+$ with centroid (width) of 0.126 (0.006)~MeV,
the $2_4^+$ at 2.144 (0.141)~MeV, and the $3_2^-$ at 2.734
(0.141)~MeV. There are also states with an uncertainly assigned $4^-$
and $4^+$ $J^\pi$, one with $J \ge 1$ and no assigned parity, and one
with no assignment.  All might influence $\alpha$-${}^6$He scattering
cross sections in the low c.m. energy range.  From
Fig.~\ref{Figure-1}, it is evident that MCAS gives a rather rich
spectrum above the particle emission thresholds, dominantly of
negative parity resonances; states that have not been observed.  The
MCAS results for the two known resonances give very good matches for
the energy centroids. Regarding widths, the $2_3^+$ result is
extremely close to that observed, but the calculated width of the
$3^-_2$ resonance is is less than half of that observed, and that for
the $4^+$ is much smaller than observed. The calculated width of the
$2^-_1$ resonance is very near that of the $2^+_4$ resonance observed
very near the same energy. Thus with this MCAS model, while we expect
some effects in cross-section evaluations due to these resonances,
there is much to be uncertain about regarding how well the calculated
spectrum matches the physical states at these energies.  Much depends
on how the resonances relate to the $\alpha$ scattering from the
ground state of ${}^6$He.

\subsection{Differential cross sections at fixed energies}
\begin{figure}[h]
\scalebox{0.7}{\includegraphics*{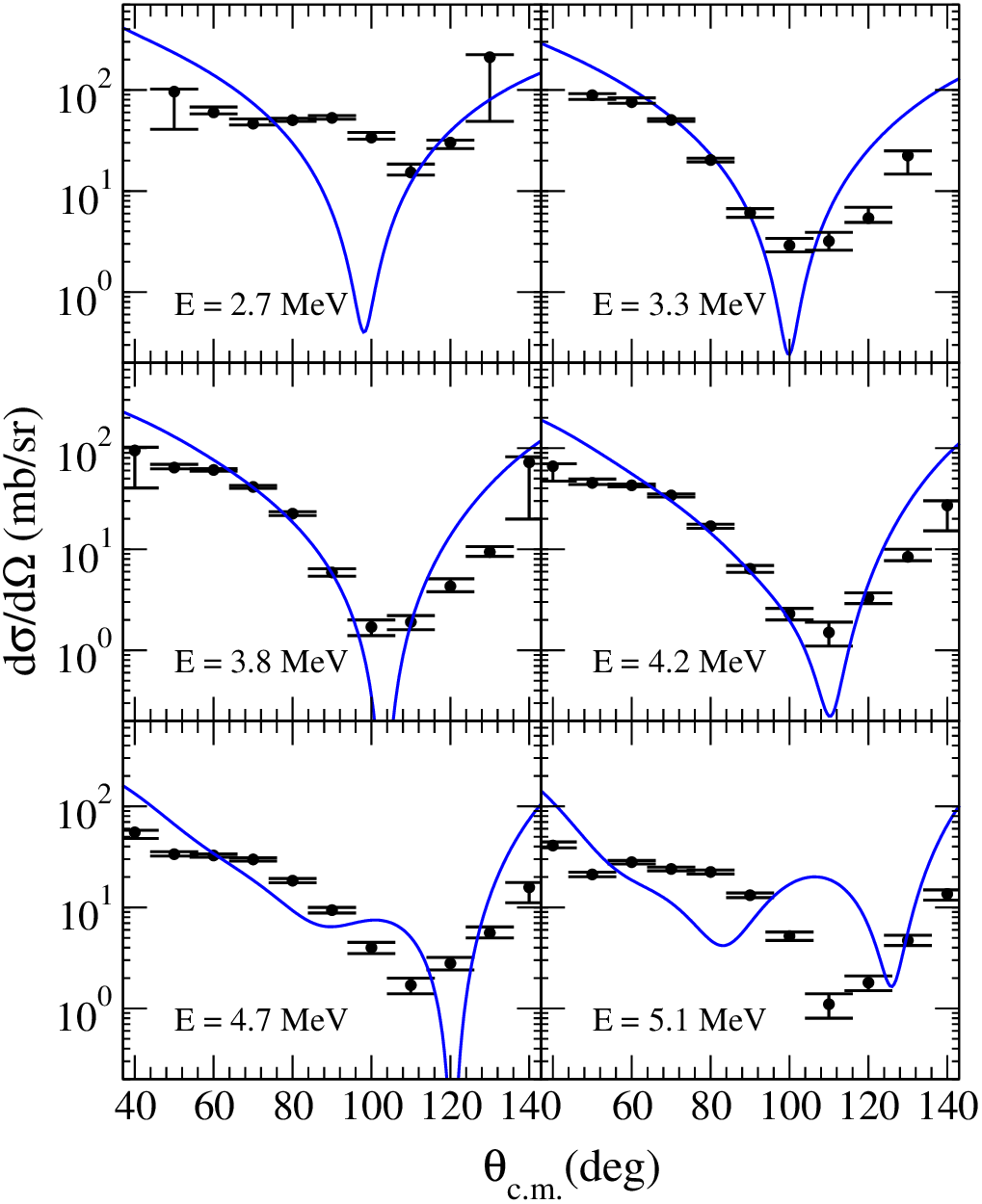}}
\caption{\label{Figure-2} (Color online) Angular distributions at
  various fixed energies as listed compared with data~\cite{Su13}.}
\end{figure}

In Fig.~\ref{Figure-2}, angular distribution data~\cite{Su13} taken at
six energies are compared with MCAS results.  We have used data
uncertainties as listed in the tabulations of the experimental
results~\cite{NNDC}.  The energies at which each of the data sets have
been taken and at which each of the MCAS evaluations were made are
indicated in the figure.
With the exception of the 2.7~MeV result, the calculated cross
sections are in reasonable agreement with the data, having appropriate
magnitudes and tracking the shape of the cross section data.  The
2.7~MeV data are distinctly different to the other energy data sets
(and of the MCAS expectation) and the known $2^+_4$ resonance in
${}^{10}$Be (2.12~MeV) has a marked effect.  MCAS finds state with the
correct angular momentum and width at an energy close to that
observed, but with incorrect parity. As that resonance may not have
the same shape as a function of energy as that observed, the
calculated 2.7 MeV cross section result is then not surprisingly a
poor representation of measured data. Further, in the 5.1~MeV result,
and arguably the 4.7~MeV result, there are calculated minima and
maxima not observed in the data.

\subsection{Energy variations at fixed scattering angles}

\begin{figure}[h]
\scalebox{0.7}{\includegraphics*{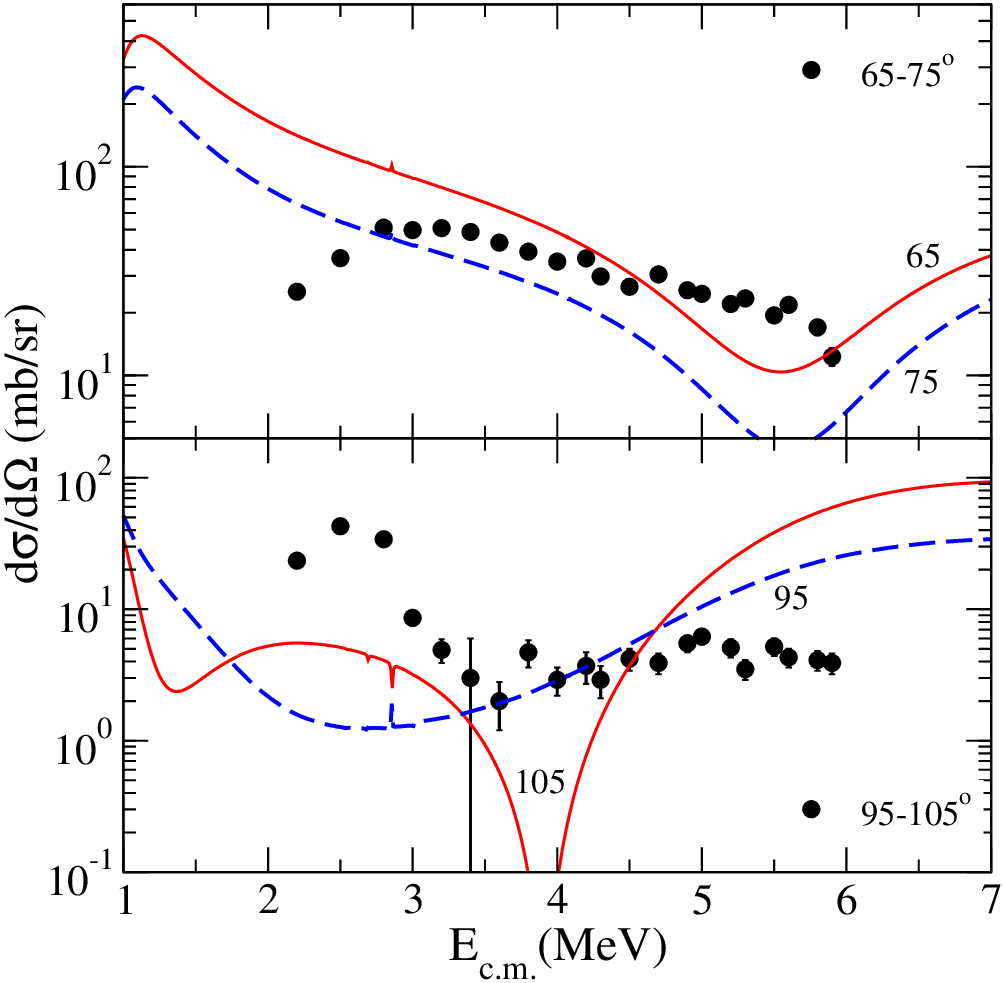}}
\caption{\label{Figure-3} (Color online) Energy variation of cross
  sections at the indicated scattering angles compared with data
  accumulated over 10 degree intervals~\cite{Su13}.}
\end{figure}

Data for energy variations of cross sections measured in two ranges of
scattering angles, $65$-$75^\circ$ and $95$-$105^\circ$, are shown in
Fig.~\ref{Figure-3}, compared to calculated MCAS results.  In the top
panel, it is shown that the 65$^\circ$ result overestimates the cross
section until $\sim$4.5~MeV, after which it underestimates it.  The
75$^\circ$ result is lower than all but the two lowest-energy data
points. Both show a dip at $\sim$5.5~MeV not seen in the data, but are
otherwise as featureless as the observed cross section. In the bottom
panel, the curves depict the MCAS results found for scattering angles
of 95$^\circ$ (solid) and 105$^\circ$ (dashed).  These are results
spanning the region of the minima of the differential cross sections;
this is revealed by the marked difference in their shapes.  Save for
the resonance-like effect in data in the 2-3~MeV range, the 95$^\circ$
result is a reasonable representation of the data, considering that
this concerns minima in the cross-sections (as observed in the fixed
energy data). The 105$^\circ$ result contains more minima and maxima
than observed, and at most energies is a poor match to data.

\subsection{Partial wave terms in the coupled-channels evaluations}

\begin{figure}[h]
\scalebox{0.7}{\includegraphics*{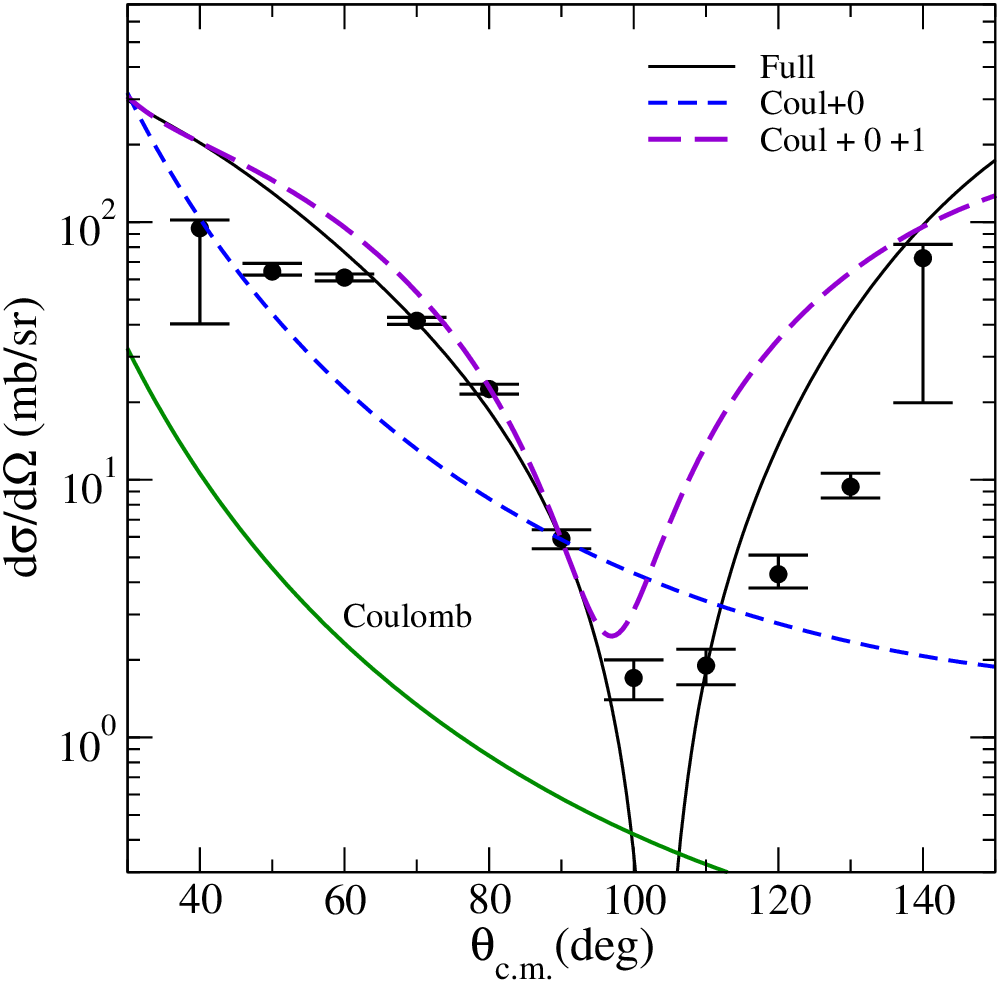}}
\caption{\label{Figure-4} (Color online) Partial wave contributions in
  the differential cross section measured at 3.8~MeV.}
\end{figure}

In Fig.~\ref{Figure-4}, the cross section taken at 3.8~MeV is compared
with individual components from the MCAS evaluation. As indicated, the
pure Coulomb cross section for the scattering of the two 3pF charge
distributions is shown by the solid curve. When the calculated
$s$-wave scattering amplitudes are added to that, the cross section
shown by the long-dashed curve results.  Adding the $p$-wave
scattering amplitudes gives the cross section depicted by the
small-dashed curve; and which features a minima near 100$^\circ$. This
is observed also in full calculated result with $d$- and $f$-waves
included (solid curve), though it is deeper.
Clearly the $p$-wave scattering arising from the couple channel
calculations define the increase in cross section values at backward
scattering angles.

\section{Conclusions}
\label{conclus}

The MCAS method has been used to solve coupled sets of
Lippmann-Schwinger equations for the $\alpha$+${}^6$He cluster,
finding a model spectrum for ${}^{10}$Be in good agreement with the
known one to more than 10~MeV excitation. A collective model was used
to define the input matrix of interaction potentials for the
coupled-channel problem in which three positive-parity states in
${}^6$He were involved.  Two of those states were taken to be
resonances themselves, with widths as defined in a data
tabulation~\cite{Ti02}. The effect of the Pauli principle on the
relative motion of the $\alpha$ and ${}^6$He was taken into account
using the OPP method. Pairing was accounted by using a monopole
interaction between the $\alpha$ and ${}^6$He.  That was found to be
necessary to give the known splitting between the ground state of the
compound system, ${}^{10}$Be, and its excited states, notably the
$2_1^+$ and $0_2^+$.

Applying MCAS to specify scattering amplitudes for positive centre of
mass energies gave angular distributions in reasonable agreement with
measured data at most energies for which the shape and magnitude
reflect a non-resonant character.  The energy variation results are
more diverse. For most energies, data taken in the range $65-75^\circ$
is enveloped by cross sections evaluated at the extremes of the range,
and the calculated results are almost as featureless as the data.  For
angles near where the minimum is observed in fixed-energy data, the
$95-105^\circ$ region, the results are of lower quality.

Overall, MCAS had produced a good representation of the ${}^{10}$Be
spectrum (treated as an $\alpha$+${}^6$He cluster), with calculated
state energies found to within a few keV of those observed across an 8
MeV interval. Calculated cross sections are a credible to good
recreation of the available data.

\begin{acknowledgments}
SK acknowledges support from the National Research Foundation of South
Africa.
\end{acknowledgments}

\bibliography{Alpha-6He}

\appendix

\section{$\alpha$-${}^6$He Coulomb potentials from two charge distributions}
\label{append}

Three different forms for Coulomb potentials have been investigated,
assuming that
\begin{enumerate}
\item both the $\alpha$ and ${}^6$He are point charge particles (each
  with charge $2e$),
\item that the ${}^6$He only had a three parameter Fermi (3pF) charge
  distribution, and
\item that both have 3pF charge distributions.
\end{enumerate}

\begin{figure}[h]
\scalebox{0.7}{\includegraphics*{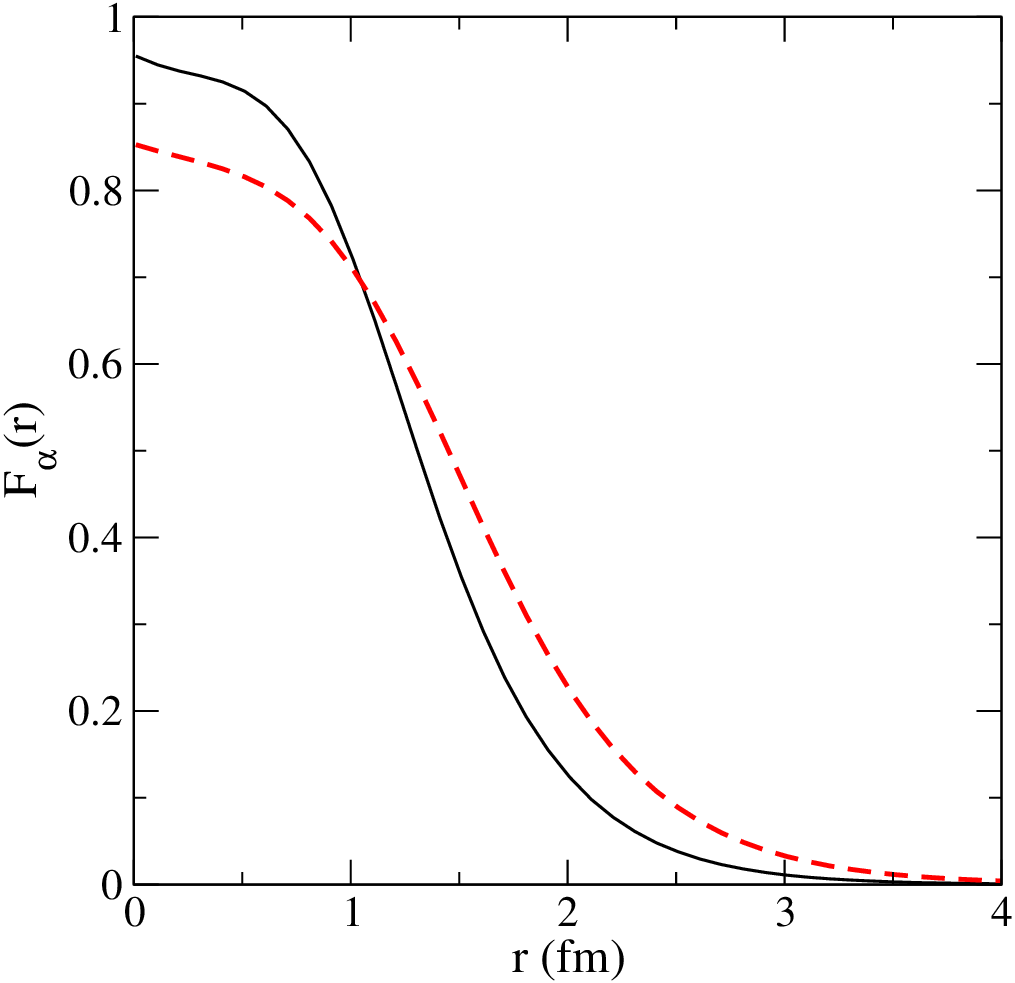}}
\caption{\label{Figure-5} (Color online)
The three parameter charge distribution for an $\alpha$ nucleus.}
\end{figure}

The 3pF charge distribution is of the form given in Eq.~(\ref{3pF-dist}), namely
\begin{equation}
\rho_{ch}(r) = \rho_0 \frac{1 + w_c \left( \frac{r}{R_c} \right)^2}
{1 + \exp\left(\frac{r - R_c}{a_c} \right)}.
\end{equation}
For an $\alpha$ particle, the parameters were determined by a fit to
the electron scattering form factor at low momentum.  Those
values~\cite{Ja74}, $R_c = 1.008$ fm, $a_c = 0.327$ fm, and
$w_c~=~0.445$, gave a charge rms radius of $R_{\rm rms}^{(c)} = 1.7$
fm.  The associated (unnormalised) charge distribution is shown by the
solid curve in Fig.~\ref{Figure-5}. The normalisation required is
$\rho_0 = 0.119$.

For ${}^6$He, it is now known~\cite{Wa04} via a laser spectroscopic
determination that its $R_{\rm rms}^{(c)} = 2.054$~fm. That result
alone does not settle the actual charge distribution, but a range of
values for a 3pF distribution that gives this value of $R_{\rm
  rms}^{(c)}$ is listed in Table~\ref{He6-3pF}.
\begin{table}[h]  
\begin{ruledtabular}
\caption{\label{He6-3pF}
Values of the parameters of 3pF charge distributions that have
$R_{\rm rms}^{(c)} \sim 2.054$ fm.}
\begin{tabular}{cc|cc|cc}
ID & $R_c$ & $a_c$ & $w$ & $a_c$ & $w$ \\
\hline
1 & 1.20 & 0.38 & 0.64 & 0.40 & 0.32 \\
2 & 1.25 & 0.38 & 0.59 & 0.40 & 0.32 \\
3 & 1.30 & 0.38 & 0.55 & 0.40 & 0.31 \\
4 & 1.35 & 0.38 & 0.51 & 0.39 & 0.38 \\
5 & 1.40 & 0.38 & 0.46 & 0.39 & 0.35 \\
6 & 1.50 & 0.37 & 0.50 & 0.38 & 0.37 \\
7 & 1.55 & 0.36 & 0.60 & 0.38 & 0.32 \\
8 & 1.60 & 0.36 & 0.51 & 0.37 & 0.38 \\
9 & 1.65 & 0.36 & 0.44 & 0.37 & 0.32 \\
\end{tabular}
\end{ruledtabular}
\end{table}
We choose the parameter set, $R_c = 1.30 $ fm, $a_c = 0.40$ fm, and
$w_c = 0.31$ as the basic set for ${}^6$He, and for those, the charge
distribution is shown by the dashed curve in Fig.~\ref{Figure-5}. The
central density required with this is $\rho_0 = 0.0704$.

With the (normalised) charge distributions, Coulomb potentials were
obtained for each of the following three cases.  They are
\begin{enumerate}
\item Case of two point charge particles:\\
The Coulomb potentials generated for this model 
is given simply by
\begin{equation}
V_{coul}(r) = \frac{4e^2}{r}.
\end{equation}
It is shown in Fig.~\ref{Figure-6} by the dot-dashed curve.
\begin{figure}[h]
\scalebox{0.7}{\includegraphics*{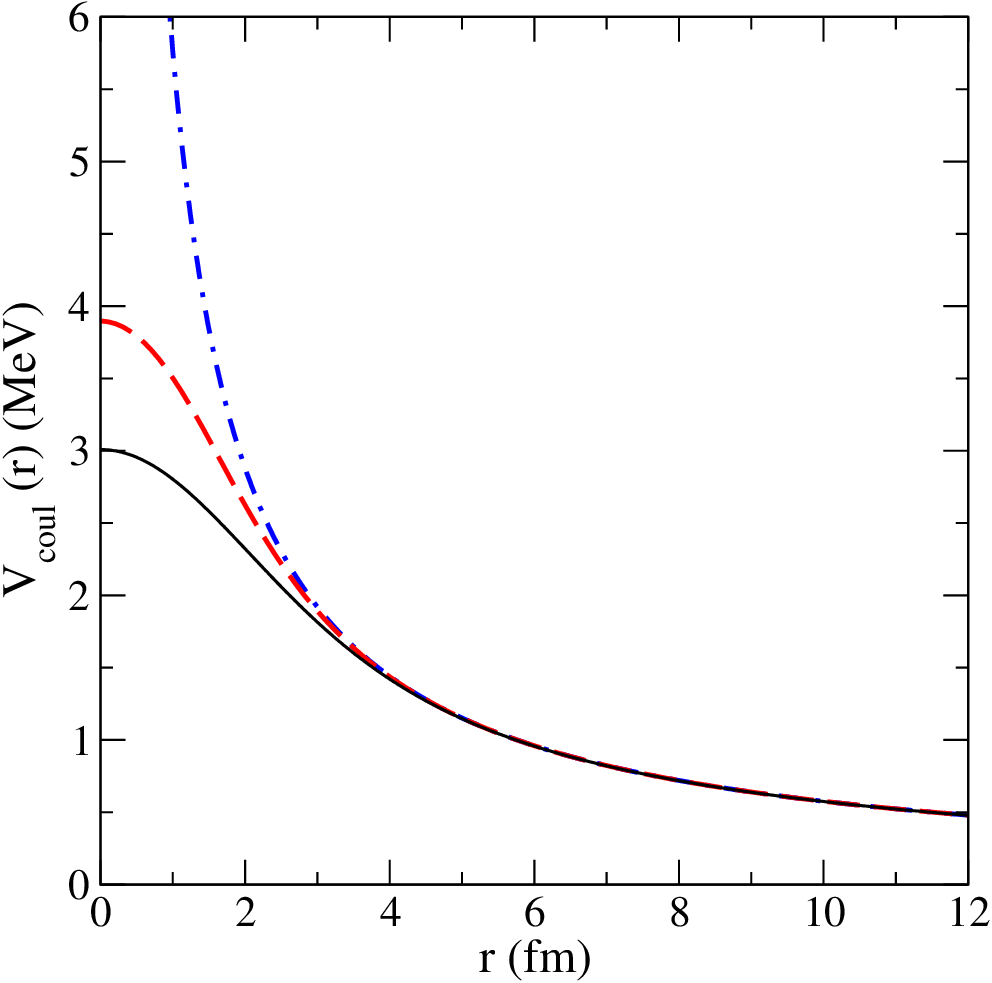}}
\caption{\label{Figure-6} (Color online)
The Coulomb potentials for each of the three $\alpha$-${}^6$He 
interaction models.}
\end{figure}
\item Case where ${}^6$He alone is given by a 3pF charge
  distribution:\\ If a nucleus has a spherical charge distribution,
  $\rho({\bf r^\prime}) = \rho_0 f_{coul}(r^\prime)$, then it is
  easily shown that the charge number $Z$ is given by
\begin{equation}
Z = 4 \pi \rho_0 \int_0^\infty f_{coul}(r^\prime)\ {r^\prime}^2\ dr^\prime\ ,
\label{eq1}
\end{equation}
thus defining the central charge density as
\begin{equation}
\rho_0 = Ze/\left[4 \pi \int_0^\infty f_{coul}(r^\prime)\ {r^\prime}^2\ 
dr^\prime \right].
\label{eq2}
\end{equation}

The Coulomb interaction felt by a  positively charged  point
test particle (having charge $2e$ for a point $\alpha$-particle) is
\begin{equation}
V_{coul}(r) = (2e) \int \rho_0 f(r^\prime) 
\frac{1}{|{\mathbf r^\prime} - {\mathbf r}|}
d{\mathbf r^\prime}
\end{equation}
and expanding in multipoles, the angular integration leaves only the
$s$-wave ($\ell = 0$) component whence
\begin{equation}
V_{coul}(r) = 2e (4\pi) \rho_0 \int_0^\infty F(r^\prime) 
v_{\ell = 0}(r^\prime, r) {r^\prime}^2 dr^\prime .
\end{equation}
where $v_{\ell =0}(r^\prime, r) = 1/r_>$ where 
$r_>$ being the greater of $r^\prime$ and $r$.
Then the radial integration splits into two terms to give
\begin{equation}
V_{coul}(r) = 4 \pi (2e) \rho_0 
\left[ \frac{1}{r} \int_0^r f_{coul}(s)\ s^2\ ds\ +\ 
\int_r^\infty \frac{1}{s} f_{coul}(s)\ s^2\ ds \right]\ 
(={\cal V}_\alpha).
\label{pt-rho}
\end{equation}
The Coulomb potential for this point-$\alpha$ interacting with ${}^6$He
with the selected 3pF charge distribution is depicted
by the dashed curve in Fig.~\ref{Figure-6}.
\item
Both $\alpha$ and ${}^6$He given by 3pF charge distributions.\\
For this case we fold the field given in Eq.(\ref{pt-rho}),
(${\cal V}_\alpha(s) $), with $\delta e$ replacing the 
$(2e)$, with the 3pF charge distribution for the second $\alpha$.
The geometry is as shown in Fig.~\ref{Figure-7}
\begin{figure}[h]
\scalebox{0.7}{\includegraphics*{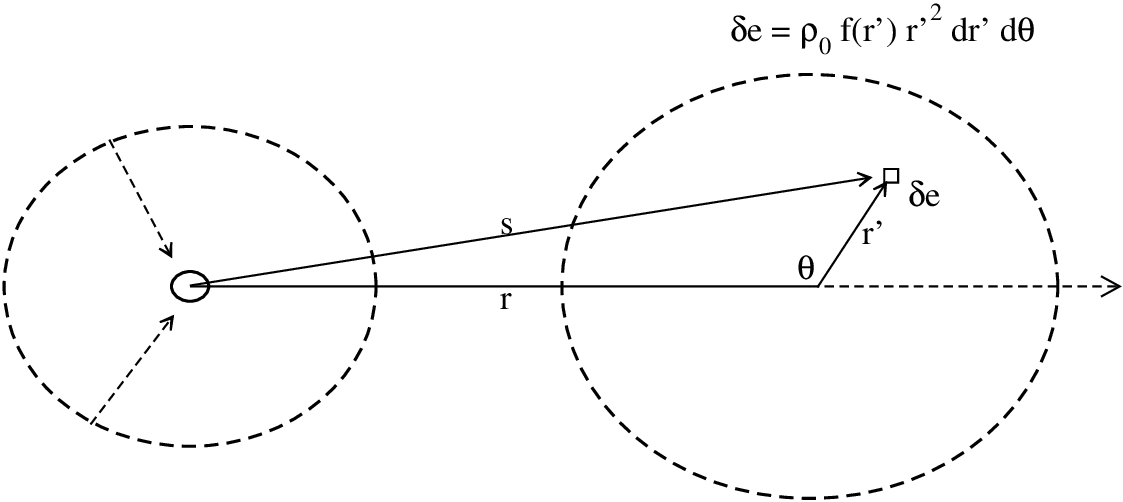}}
\caption{\label{Figure-7} (Color online)
The geometry for both $\alpha$ nuclei having a 3pF charge distribution.}
\end{figure}
We seek a result for use in MCAS of $V_{coul}(r)$ which,
with $s = \sqrt{r^2 + {r'}^2 - 2 s r^\prime \cos(\theta })$,
 is given by
\begin{equation}
V_{coul}(r) = 2\pi \int_0^\infty {r^\prime}^2 f(r^\prime)
\int_0^\pi {\cal V}_\alpha(s)\ sin(\theta)\ d\theta .
\end{equation}
The result is shown by the solid curve in Fig.~\ref{Figure-6}.
\end{enumerate}

\end{document}